\documentclass[twocolumn,aps,prl,showpacs,superscriptaddress,,amssymb]{revtex4}
\usepackage{graphicx}
\usepackage{dcolumn}
\usepackage{bm}
\usepackage{amsmath}
\usepackage{relsize}
\usepackage{xspace}
\def\babar{\mbox{\slshape B\kern-0.1em{\smaller A}\kern-0.1em
    B\kern-0.1em{\smaller A\kern-0.2em R}}}
\def\pep2{PEP-II}

\RequirePackage{xspace}

\usepackage{relsize}
\def\babar{\mbox{\slshape B\kern-0.1em{\smaller A}\kern-0.1em
    B\kern-0.1em{\smaller A\kern-0.2em R}}}

\def\epem       {\ensuremath{e^+e^-}\xspace}

\def\mumu       {\ensuremath{\mu^+\mu^-}\xspace}

\def\ellell     {\ensuremath{\ell^+ \ell^-}\xspace}

\def\q     {\ensuremath{q}\xspace}

\def\ccbar {\ensuremath{c\overline c}\xspace}

\def\pip   {\ensuremath{\pi^+}\xspace}

\def\pipi  {\ensuremath{\pi^+\pi^-}\xspace}

\def\Kbar  {\kern 0.2em\overline{\kern -0.2em K}{}\xspace}

\def\Kz    {\ensuremath{K^0}\xspace}
\def\Kzb   {\ensuremath{\Kbar^0}\xspace}
\def\KzKzb {\ensuremath{\Kz \kern -0.16em \Kzb}\xspace}
\def\Kp    {\ensuremath{K^+}\xspace}
\def\Km    {\ensuremath{K^-}\xspace}

\def\KpKm  {\ensuremath{\Kp \kern -0.16em \Km}\xspace}
\def\KS    {\ensuremath{K^0_{\scriptscriptstyle S}}\xspace}

\def\Dbar    {\kern 0.2em\overline{\kern -0.2em D}{}\xspace}

\def\Dz      {\ensuremath{D^0}\xspace}
\def\Dzb     {\ensuremath{\Dbar^0}\xspace}
\def\DzDzb   {\ensuremath{\Dz {\kern -0.16em \Dzb}}\xspace}
\def\Dp      {\ensuremath{D^+}\xspace}
\def\Dm      {\ensuremath{D^-}\xspace}

\def\DpDm    {\ensuremath{\Dp {\kern -0.16em \Dm}}\xspace}

\def\B       {\ensuremath{B}\xspace}
\def\Bbar    {\kern 0.18em\overline{\kern -0.18em B}{}\xspace}

\def\Bz      {\ensuremath{B^0}\xspace}
\def\Bzb     {\ensuremath{\Bbar^0}\xspace}
\def\BzBzb   {\ensuremath{\Bz {\kern -0.16em \Bzb}}\xspace}
\def\Bu      {\ensuremath{B^+}\xspace}
\def\Bub     {\ensuremath{B^-}\xspace}
\def\Bp      {\ensuremath{\Bu}\xspace}

\def\BpBm    {\ensuremath{\Bu {\kern -0.16em \Bub}}\xspace}

\def\BorBbar    {\kern 0.18em\optbar{\kern -0.18em B}{}\xspace}
\def\DorDbar    {\kern 0.18em\optbar{\kern -0.18em D}{}\xspace}
\def\KorKbar    {\kern 0.18em\optbar{\kern -0.18em K}{}\xspace}

\def\jpsi     {\ensuremath{{J\mskip -3mu/\mskip -2mu\psi\mskip 2mu}}\xspace}

\mathchardef\Upsilon="7107
\def\Y#1S{\ensuremath{\Upsilon{(#1S)}}\xspace}

\def\FourS {\Y4S}

\mathchardef\Deltares="7101
\mathchardef\Xi="7104
\mathchardef\Lambda="7103
\mathchardef\Sigma="7106
\mathchardef\Omega="710A

\def\Deltabar{\kern 0.25em\overline{\kern -0.25em \Deltares}{}\xspace}
\def\Lbar{\kern 0.2em\overline{\kern -0.2em\Lambda\kern 0.05em}\kern-0.05em{}\xspace}
\def\Sigbar{\kern 0.2em\overline{\kern -0.2em \Sigma}{}\xspace}
\def\Xibar{\kern 0.2em\overline{\kern -0.2em \Xi}{}\xspace}
\def\Obar{\kern 0.2em\overline{\kern -0.2em \Omega}{}\xspace}
\def\Nbar{\kern 0.2em\overline{\kern -0.2em N}{}\xspace}
\def\Xb{\kern 0.2em\overline{\kern -0.2em X}{}\xspace}

\def\pt         {\mbox{$p_T$}\xspace}
\def\mes        {\mbox{$m_{\rm ES}$}\xspace}

\def\DeltaE     {\mbox{$\Delta E$}\xspace}

\newcommand{\tev}{\ensuremath{\mathrm{\,Te\kern -0.1em V}}\xspace}
\newcommand{\gev}{\ensuremath{\mathrm{\,Ge\kern -0.1em V}}\xspace}
\newcommand{\mev}{\ensuremath{\mathrm{\,Me\kern -0.1em V}}\xspace}
\newcommand{\kev}{\ensuremath{\mathrm{\,ke\kern -0.1em V}}\xspace}
\newcommand{\ev}{\ensuremath{\mathrm{\,e\kern -0.1em V}}\xspace}
\newcommand{\gevc}{\ensuremath{{\mathrm{\,Ge\kern -0.1em V\!/}c}}\xspace}
\newcommand{\mevc}{\ensuremath{{\mathrm{\,Me\kern -0.1em V\!/}c}}\xspace}
\newcommand{\gevcc}{\ensuremath{{\mathrm{\,Ge\kern -0.1em V\!/}c^2}}\xspace}
\newcommand{\mevcc}{\ensuremath{{\mathrm{\,Me\kern -0.1em V\!/}c^2}}\xspace}

\def\cm   {\ensuremath{\rm \,cm}\xspace}

\def\mm   {\ensuremath{\rm \,mm}\xspace}

\def\invfb   {\ensuremath{\mbox{\,fb}^{-1}}\xspace}

\def\mus  {\ensuremath{\rm \,\mus}\xspace}

\def\mus        {\ensuremath{\,\mu{\rm s}}\xspace}    

\def\rad{\ensuremath{\rm \,rad}\xspace}

\def\to                 {\ensuremath{\rightarrow}\xspace}

\def\pep2{PEP-II}

\def\gsim{{~\raise.15em\hbox{$>$}\kern-.85em
          \lower.35em\hbox{$\sim$}~}\xspace}
\def\lsim{{~\raise.15em\hbox{$<$}\kern-.85em
          \lower.35em\hbox{$\sim$}~}\xspace}

\xspace

\newcommand{\jprlBase}       {Phys.\ Rev.\ Lett.\xspace}
\newcommand{\jprBase}        {Phys.\ Rev.\xspace}
\newcommand{\jplBase}        {Phys.\ Lett.\xspace}
\newcommand{\nimBaseA}       {Nucl.\ Instr.\ Meth.\xspace}

\newcommand{\npBase}         {Nucl.\ Phys.\xspace}
\newcommand{\zpBase}         {Z.\ Phys.\xspace}

\newcommand{\mpl}       [1]  {{Mod.\ Phys.\ Lett.\ {\bf #1}}}

\newcommand{\nima}      [1]  {\nimBaseA~A~{\bf #1}}

\newcommand{\npb}       [1]  {\npBase\ B~{\bf #1}}

\newcommand{\plb}       [1]  {\jplBase\ B~{\bf #1}}

\newcommand{\jprl}      [1]  {\jprlBase\ {\bf #1}}
\newcommand{\jprd}      [1]  {\jprBase\ D~{\bf #1}}

\newcommand{\zpc}       [1]  {\zpBase\ C~{\bf #1}}

\def\jetset74   {\mbox{\tt Jetset \hspace{-0.5em}7.\hspace{-0.2em}4}\xspace}

\newcommand{\bzjpsiks}{\ensuremath{B^0 \to \jpsi\KS }}
\newcommand{\bpjpsikp}{\ensuremath{B^+ \to \jpsi K^+ }}

\begin{document}
\title{\boldmath
Measurement of the \Bu/\Bz Production Ratio from the \FourS\ Meson using
\bpjpsikp\ and \bzjpsiks\ Decays}

\author{B.~Aubert}
\author{R.~Barate}
\author{D.~Boutigny}
\author{F.~Couderc}
\author{J.-M.~Gaillard}
\author{A.~Hicheur}
\author{Y.~Karyotakis}
\author{J.~P.~Lees}
\author{V.~Tisserand}
\author{A.~Zghiche}
\affiliation{Laboratoire de Physique des Particules, F-74941 Annecy-le-Vieux, France }
% ...
\collaboration{The \babar\ Collaboration}
\noaffiliation

\date{\today}
 \begin{abstract}
We report a measurement of the production ratio of charged and neutral $B$ mesons
from \FourS decays based on the ratio of efficiency-corrected yields for the charmonium 
modes $\jpsi K^+$ and $\jpsi \KS$ with 81.9\invfb of data collected with the \babar\ detector on the \FourS resonance at 10.580 \gev.
We find a value of $1.006\pm 0.036 (stat)\pm 0.031 (sys)$ for the ratio
$R^{+/0}=\Gamma(\FourS\to\Bu\B^{-})/\Gamma(\FourS\to\Bz\Bzb)$.
\end{abstract}
\maketitle

A measurement of the $B^+/B^0$ production ratio
$$R^{+/0}=\frac{\Gamma(\FourS\to\Bu\B^{-})}{\Gamma(\FourS\to\Bz\Bzb)}$$ 
from the \FourS meson is an essential element in determining branching fractions and 
quark-mixing matrix elements at the \B factory experiments. It can also 
provide information about the structure of the \FourS meson that can be used
to discriminate between available models. 

Over the past 15 years it has been frequently assumed that $R^{+/0}$ is equal to one, 
although many models predict that this may not be the case.
Early calculations predicted that the ratio could be up to 20\% greater than one, due to large Coulomb corrections~\cite{atwood}. 
Taking into account the structure of the \B and \FourS reduces the effect of the Coulomb interaction 
and can even lead to the ratio being less than unity~\cite{lepage}. 
With the prospect of precision measurements from the \B factories, there has been a recent revival in theoretical work on the subject. 
A more detailed calculation has been done in a non-relativistic effective field theory with
$B^*$ intermediate states in the pion potential, which introduces isospin-breaking in strong interactions. These calculations predict a value
$1.1-1.2$~\cite{manohar}.  
Other calculations attempting to take into account the structure of the mesons and hadronic final state interactions predict a ratio $0.9-1.2$~\cite{voloshin},
but with rapid variation as a function of the center-of-mass energy near the \FourS
resonance. However, such rapid variation in the charged-to-neutral ratio
has not been seen in scans across the $\phi(1020)$ resonance~\cite{phi}.
For the \FourS, there are published
measurements of  $R^{+/0}$ by CLEO ($1.04\pm 0.07\pm 0.04$~\cite{CLEO1},
$1.058\pm 0.084\pm 0.136$~\cite{CLEO2}), \babar\ ($1.10\pm 0.06\pm 0.05$~\cite{BABAR1} with
20\invfb), and Belle ($1.01\pm 0.03\pm 0.09$~\cite{BELLE} with 29\invfb). Now that a significantly larger \FourS data sample is available at \babar\
we can reduce the statistical uncertainty to the point where it is possible to confront the various theoretical
predictions.

In this analysis we use the decay modes \bzjpsiks\ and \bpjpsikp~\cite{cc}, where \jpsi\to\ellell and \KS \to \pipi,
to measure the \Bu/\Bz production ratio.
These decays are good candidates for measuring $R^{+/0}$ 
since isospin violation in the $\B\to\jpsi K$ decays is expected to be small in the Standard Model, of order $\lambda^3\thickapprox0.01$~\cite{fleischer} when
rescattering is small, where 
$\lambda$ is defined as the sine of the Cabibbo angle.  

The data used in this analysis were collected with the \babar\ detector at 
the \pep2\ $e^+e^-$ storage ring.
The data sample corresponds to 81.9 \invfb of integrated luminosity 
collected at the \FourS\ resonance.
The distribution of center-of-mass energies due to the beam-energy spread is Gaussian with $\sigma = 4.6\mev$~\cite{beamspread}. The mean energy of
our sample is 10.580\gev, with all data accumulated within one sigma of this value.

      The \babar\ detector is fully described elsewhere~\cite{babar}. It consists 
of a charged-particle tracking
system, a Cherenkov detector (DIRC) for particle identification, an electromagnetic 
calorimeter, and a system for muon
identification. The tracking system consists of a 5-layer, double-sided 
silicon vertex tracker and a
40-layer drift chamber (filled with a mixture of helium and isobutane), both in a 
1.5-T magnetic field supplied by a
superconducting solenoidal magnet. The DIRC is an imaging Cherenkov detector 
relying on total internal
reflection in the radiator. The electromagnetic calorimeter consists of 6580 
CsI(Tl) crystals. The iron flux return
is segmented and instrumented with resistive plate chambers for muon identification.

Hadronic events are selected by requiring the presence of at least three tracks 
in the angular region $0.41<\theta_{\rm LAB}<2.54 {\rm \rad}$, where $\theta_{\rm LAB}$ is the polar angle with respect to the beam direction. The ratio 
between the 2nd and 0th order Fox-Wolfram~\cite{FOX} moments must be less than 0.5.
We also require that the total energy of all particles in the event be
greater than 4.5\gev. The primary vertex, which is constructed from charged tracks with impact parameter less than 1mm in the 
plane transverse to the beam direction, 
must be within 0.5\cm
of the beam spot in the plane transverse to the beam direction
and within 6\cm\ along the beam direction.

We reconstruct candidates for \jpsi\ mesons in the decay modes $\jpsi\to\epem$ 
and $\mu^+\mu^-$. For $\jpsi\to\epem$ decays one track is required to 
pass a tight electron selection and the other a loose requirement~\cite{BABAR1}, 
while for $\jpsi\to\mu^+\mu^-$ decays we require one track to pass a loose muon selection and the other
a minimum-ionizing requirement~\cite{BABAR1}. The daughter tracks of the \jpsi candidate are 
required to have 12 hits in the drift chamber, lie in the angular range $0.41<\theta_{\rm LAB}<2.409 {\rm \rad}$ 
for electrons and  $0.41<\theta_{\rm LAB}<2.54 {\rm \rad}$ for muons, 
and have a transverse momentum of at least 100\mevc.  
To increase the efficiency of the event selection, electron candidate tracks
are combined with photon candidates to recover some of the energy lost in bremsstrahlung~\cite{BABAR1}.
A geometric vertex constraint fit is applied
to the lepton track pair.
The invariant mass requirements
for the \jpsi\to\epem and \jpsi\to\mumu channels are $2.95<M_{\epem}<3.14 \gevcc$
and $3.06<M_{\mumu}<3.14 \gevcc$. We require
$|\cos \theta_{\ell}|$ to be less than 0.8 and 0.9 for \jpsi\to\epem and \jpsi\to\mumu respectively. The helicity angle $\theta_{\ell}$
is the angle in the \jpsi rest frame between the positively charged \jpsi daughter and the reversed $K$
flight direction in the \B meson rest frame.  

We reconstruct \KS meson candidates from two charged tracks, which are not required to originate from the
interaction point or to have drift chamber hits, in contrast to the \jpsi daughters.  The tracks are assigned the pion mass to compute $M_{\pipi}$, which is
required 
to lie in the range $0.490-0.505\gevcc$. Also, in order to reject combinatorial background, we only retain 
candidates with a fitted \KS vertex displaced more than 1\mm from the \jpsi vertex.
Candidates for $K^+$ mesons are assigned the kaon mass and are required to form a vertex with the \jpsi candidate.  
No particle identification requirements are made for this track.

The selection of \B\ candidates relies on the kinematic constraints given by the \FourS\ initial state. Two largely
uncorrelated variables are used: 
the energy-substituted \B mass $\mes = \sqrt{(s/2+{\bf p_0}\cdot{\bf p_B})^2/E_0^2-p_B^2}$,
where the subscripts 0 and \B refer to the \epem system and the \B candidate respectively,
$s$ is the square of the center-of-mass energy,
and energies ($E$) and momentum vectors (${\bf p}$) are computed in the laboratory frame; and·
$\DeltaE=E_B^*-\sqrt{s}/2$, where $E_B^*$ is the \B candidate energy in the center-of-mass frame.
In cases where multiple \B\ candidates are present in the same event, 2\% of the total, only the one with the smallest absolute value of \DeltaE is retained.

       The signal region in the \mes-\DeltaE plane is defined by $5.27<\mes<5.29\gevcc$ and $\vert\DeltaE\vert<3\sigma(\DeltaE)$. The observed resolutions for
data
and Monte Carlo for the different modes are listed in Table~\protect\ref{tab:res}.  The \mes sideband is defined by $5.20<\mes<5.27\gevcc$ and
$\vert\DeltaE\vert<3\sigma(\DeltaE)$.
Upper and lower \DeltaE sidebands, used for the evaluation of systematic errors, are defined as $50< \DeltaE < 120\mev$ and $-120< \DeltaE < -50\mev$.

      Since we are measuring the ratio of \Bu to \Bz efficiency-corrected yields
many of the selection requirements are in common and have been optimized previously~\cite{BABAR1}.
Therefore for this analysis, we have only reconsidered the optimization of the statistical uncertainty of the measurement due to
those requirements that are different for the two modes.
These requirements include the \KS flight length, \KS mass window, and the \DeltaE window.
The optimization of these variables maximizes the ratio $N_{\rm cand}/\sqrt{\sigma^2(N_{\rm cand})+\sigma^2(N_{\rm bkg})}$
where $\sigma(N_{\rm cand})$ and $\sigma(N_{\rm bkg})$ are the uncertainties on the
number of signal candidates $N_{\rm cand}$ predicted by Monte Carlo (MC) simulation
and combinatorial background $N_{\rm bkg}$, respectively, 
that pass the event selection procedure. 
$N_{\rm cand}$ is defined as number of events in the signal region.
      
\begin{table}[tbh]
\caption{Summary of the resolution for \DeltaE in data and MC simulation.}
\begin{tabular}{l c cc }
\multicolumn{4}{c}{}\\
\hline
\multicolumn{2}{c}{Mode} & \multicolumn{2}{c} { $\sigma(\DeltaE)$ [MeV]} \\
\hline
  \B       & \jpsi         & MC & Data \\
\hline
$\B^+$& \epem& $11.42\pm 0.11$ &  $10.87\pm 0.25$ \\
$\B^+$& \mumu&  $9.72\pm 0.07$ &  $ 9.25\pm 0.20$ \\
\hline
$\B^0$& \epem& $9.50\pm 0.11$  &  $10.02\pm 0.42$ \\
$\B^0$& \mumu& $7.92\pm 0.07$  &  $ 8.52\pm 0.32$ \\

\hline
\end{tabular}

\label{tab:res}

\end{table}

\begin{figure}
\includegraphics[width=6.6cm,height=6.6cm]{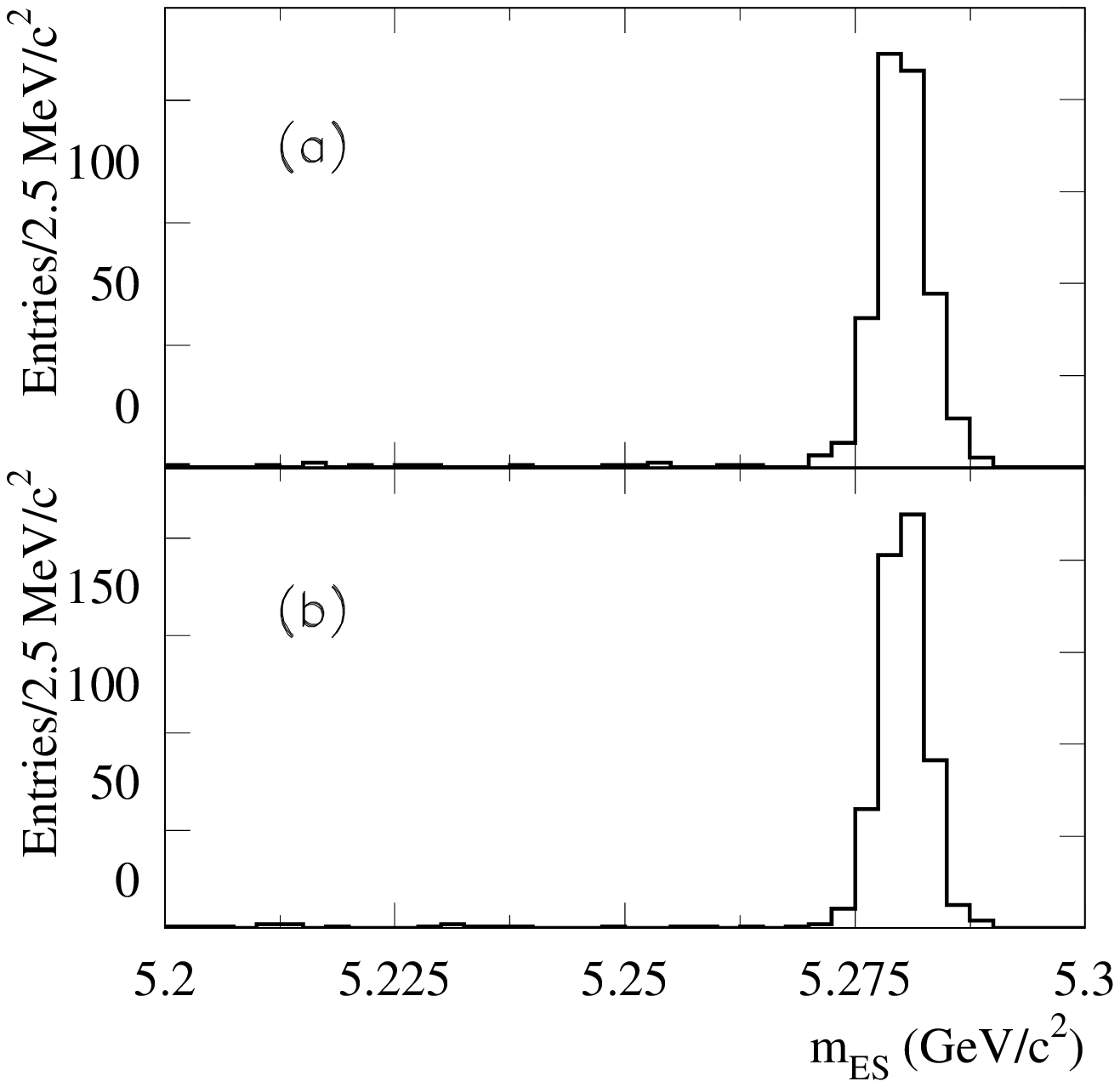}
\includegraphics[width=6.6cm,height=6.6cm]{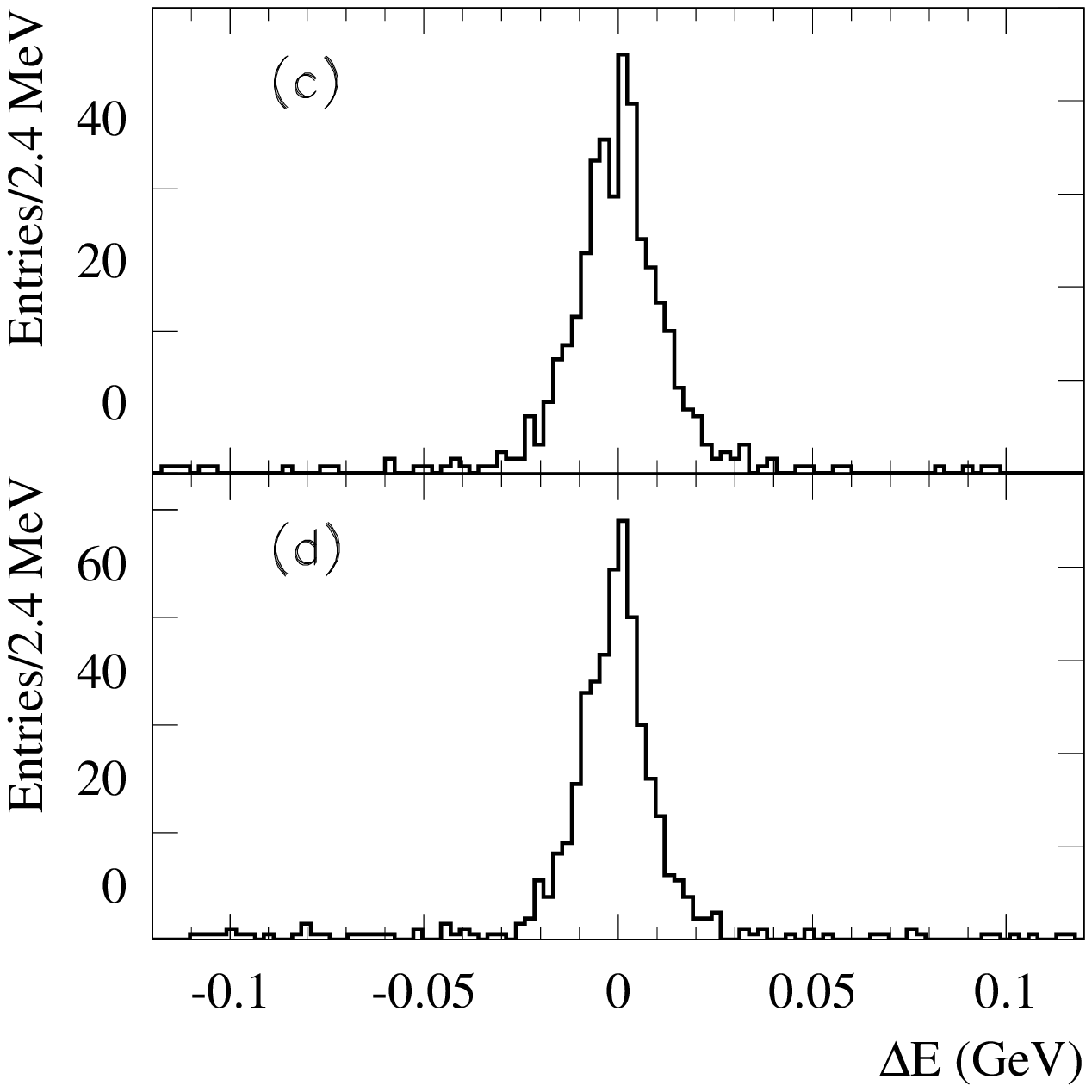}
\caption{Distribution of \mes for $|\DeltaE|<3\sigma$ in the \bzjpsiks\ sample for (a) $\jpsi\to\epem$ and (b) $\jpsi\to\mu^+\mu^-$.
Distribution of \DeltaE for $\mes>5.27\gevcc$ in the \bzjpsiks\ sample for (c) $\jpsi\to\epem$ and (d) $\jpsi\to\mu^+\mu^-$.}
\label{fig:ks}
\end{figure}

\begin{figure}
\includegraphics[width=6.6cm,height=6.6cm]{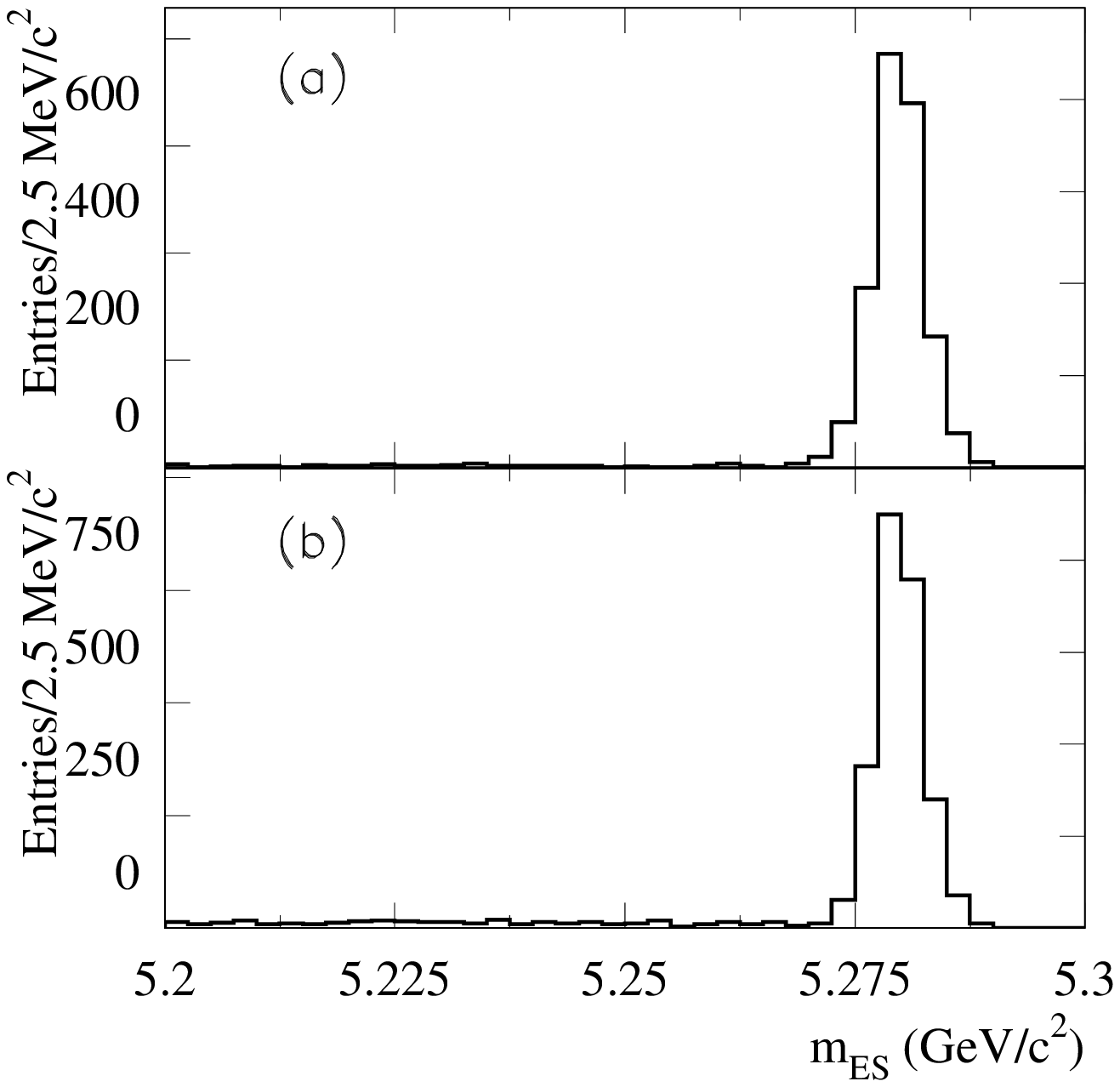}
\includegraphics[width=6.6cm,height=6.6cm]{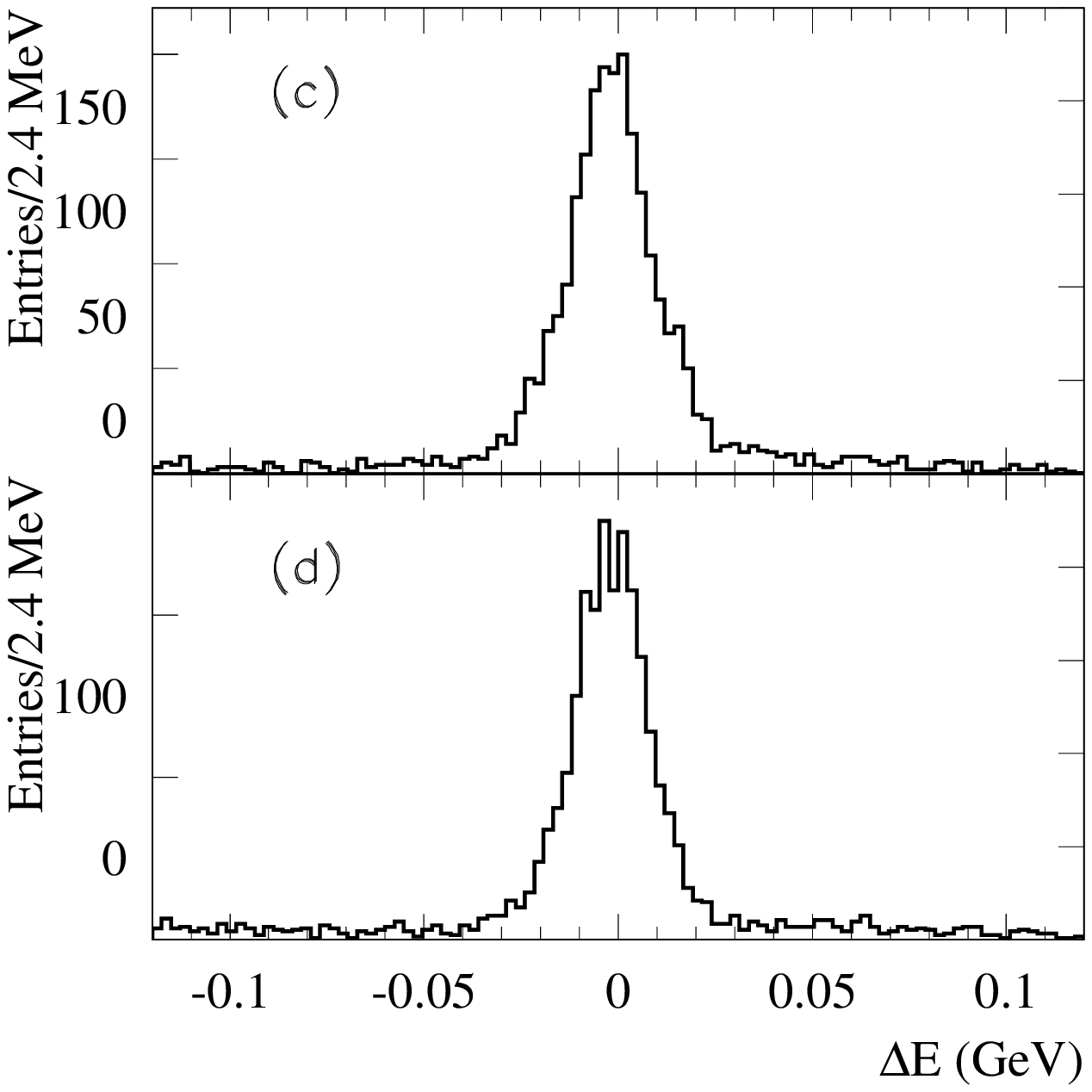}
\caption{Distribution of \mes for $|\DeltaE|<3\sigma$ in the \bpjpsikp\ sample for (a) $\jpsi\to\epem$ and (b) $\jpsi\to\mu^+\mu^-$.
Distribution of \DeltaE for $\mes>5.27\gevcc$ in the \bpjpsikp\ for (c) $\jpsi\to\epem$ and (d) $\jpsi\to\mu^+\mu^-$.}
\label{fig:kp}
\end{figure}

We fit the \mes distribution in the \mes sideband with an empirical phase-space-motivated function introduced by ARGUS~\cite{ARGUS}.  The fitted distribution
is then integrated over the signal region to determine the number of combinatorial 
background events $N_{\rm bkg}$. 
In addition to combinatorial backgrounds there are other background sources, mostly in \B decays to charmonium, that peak near the \B mass in
\mes. These peaking sources are negligible for the neutral \B sample, but include small contributions from $\Bz\to\jpsi\KS$ and $\Bu\to\jpsi\pip$ for the
charged \B sample. Requiring particle identifications on the $K^+$ candidate will reduce these contributions but introduces a larger systematic error.  
To determine the number of background events that peak in the \mes signal region, $N_{\rm peak}$, 
we use appropriately combined MC samples of continuum \epem\ and generic~\cite{generic} $B \bar{B}$ events (with signal events removed), which have been 
scaled to the integrated luminosity of the data sample.
This distribution is then fitted with an ARGUS function as described above. We determine $N_{\rm peak}$ by counting the
number of events in the signal region and subtracting the integral of the ARGUS function over this same
region. The signal yield is then defined by $N_{\rm signal}=N_{\rm cand}-N_{\rm bkg}-N_{\rm peak}$.
The observed distributions in \mes and \DeltaE for \bzjpsiks\ and \bpjpsikp\ candidates in data are shown in Figs.~\protect\ref{fig:ks} and~\protect\ref{fig:kp}
respectively.

The efficiency-corrected ratio of observed events is given by:
\begin{eqnarray*}
\frac{N_{\rm signal}^+/\epsilon_+}{N_{\rm signal}^0/(f\epsilon_0)}
&=& R^{+/0}\frac{{\cal B}(\Bu\to\jpsi\Kp)}{{\cal B}(\Bz\to\jpsi\KS)} \\
&=& R^{+/0}\frac{2\Gamma(\Bu\to\jpsi\Kp)\tau_+}{\Gamma(\Bz\to\jpsi K^{0})\tau_0} \\
\end{eqnarray*}
where $f=68.60\pm0.27\%$~\cite{PDG2002} is the \KS\to\pipi branching fraction, $\tau_{+}/\tau_{0}=1.083\pm0.017$~\cite{PDG2002}
is the ratio of \Bu and \Bz lifetimes, and $\epsilon$ is the selection efficiency.
Therefore, assuming isospin invariance in the $\B\to\jpsi K$ decay, $\Gamma(B^+\to\jpsi\Kp) = \Gamma(B^0\to\jpsi K^{0})$~\cite{phasesp}, the ratio of
efficiency-corrected yields is determined from:
\begin{equation}
R^{+/0}=\frac{N_{\rm signal}^{+}\epsilon_{0} f}{2N_{\rm signal}^{0}\epsilon_{+} }\frac{\tau_{0}}{\tau_{+}}.
\label{BR_calc}
\end{equation}
The ratio of efficiency-corrected yields is
determined separately for $\jpsi\to\epem$ and $\jpsi\to\mu^+\mu^-$ so that lepton
identification efficiencies cancel. The separate measurements are then averaged, keeping track of correlated uncertainties, to produce
a final value for $R^{+/0}$. 

Sources of systematic uncertainties can be classified into those arising from uncertainties on efficiencies and 
those from candidate selection and backgrounds.
The efficiency uncertainties are due to \KS reconstruction, tracking, and kaon/pion tracking efficiency differences. 
In the ratio of the efficiency-corrected yields, the tracking uncertainty is due to the 
extra track required to reconstruct the \bzjpsiks\ mode. 
We determine the relative kaon/pion tracking reconstruction efficiency
by comparing the ratio of efficiencies for \bpjpsikp\ and \Bp\to\jpsi\pip Monte Carlo.
The systematic error of 0.6\% is taken to be half the size of the estimated difference.
Finally, for the uncertainty on the \KS efficiency we take a sample of
inclusive \KS candidates that are binned in transverse momentum (\pt), 
laboratory polar angle ($\theta_{\rm LAB}$), and transverse flight length ($dr$).
A relative correction for reconstruction of a displaced \KS\ candidate is determined in each
\pt\ and $\theta_{\rm LAB}$ bin by assuming the tracking efficiency for a short-lived
\KS close to the interaction region is the same as for prompt tracks. Thus, the ratio of data to MC relative efficiency is normalized to unity for small $dr$ and
then used to
derive a MC correction factor for larger displacements.
By varying the size of the $dr$, \pt, and $\theta_{\rm LAB}$ bins we determine a systematic uncertainty for this procedure.
The normalization bin for the correction is well inside the radius of the beam pipe. We vary the definition of this bin as a check of
the hypothesis that these tracks have the same efficiency as normal charged tracks.

\begin{table}[!tbhp]
\vspace*{3pt}
\caption{Summary of the relative systematic uncertainties on the efficiency-corrected yields.}
\begin{tabular}{lc ccccccc}
\multicolumn{9}{c}{}\\
\hline
\multicolumn{2}{c} {Mode} &\multicolumn{7}{c}{Parameters (\%)} \\
\hline
             & &           &           &              &ARGUS    & Peaking & Vary    &\\
 \B&\jpsi        &$\epsilon_{\rm Trk}$ & $\epsilon_{\rm K^+/\pip}$ & $\epsilon_{\rm \KS}$ & Bkgd.& Bkgd.   & Selection  & Total \\
\hline
$\B^+$& \epem    &    -            & 0.6 &   -  & 0.5 &0.1 & 0.1  & 0.8 \\
$\B^+$& \mumu    &    -            & 0.6 &   -  & 0.6 &0.4 & 1.0  & 1.4 \\
\hline
$\B^0$& \epem     &  1.3           &  -  &  1.8 & 0.8 &0.1 & 0.2  & 2.4 \\
$\B^0$& \mumu     &  1.3           &  -  &  1.8 & 0.5 &0.1 & 1.3  & 2.6 \\
\hline
\end{tabular}
\label{syserr}
\end{table}

The selection and background systematic uncertainties are attributed to the selection criteria, the ARGUS background shape, and the peaking background
subtraction.
The selection requirements on the \KS mass, \KS flight distance, and \DeltaE are varied within reasonable ranges. The \KS mass and \DeltaE selection windows
were increased and decreased from the nominal value by half a sigma and the \KS vertex displacement
requirement was removed.  The largest difference from the nominal efficiency-corrected yield
is taken conservatively as a systematic uncertainty.
The continuum background systematic uncertainty is determined by varying the ARGUS parameter by two sigma to
account for any model dependence.  
The peaking background uncertainty is evaluated from the discrepancy between data and MC in the the upper and lower \DeltaE sidebands.
The larger of the two discrepancies is taken as the systematic uncertainty.  This is a more conservative approach than using the uncertainties 
for the relevant branching fractions.  
Table~\protect\ref{syserr} summarizes the sources of systematic uncertainty for this analysis.

\begin{table}[!htbp]
\vspace*{3pt}
\caption{Summary of values needed to determine the efficiency corrected yields.}
\begin{tabular}{lccccc}
\multicolumn{6}{c}{}\\
\hline
 \multicolumn{2}{c}{Mode}      &      \multicolumn{4}{c}{Parameters}\\
\hline
  \B& \jpsi &$N_{\rm cand}$   & $N_{\rm bkg}$ & $N_{\rm peak}$  & Efficiency (\%)            \\
\vspace*{-10pt} & \\
\hline
 $B^+$& $e^+  e^-$ &  2213 &    19.5 $\pm$     5.0 &     9.6 $\pm$     3.2 & 40.8 $\pm$ 0.4 \\
 $B^0$& $ e^+  e^-$ &   502 &     2.6 $\pm$     2.0 &     2.4 $\pm$     1.5 & 29.9 $\pm$ 0.4\\
\hline 
$B^+$ &$\mu^+\mu^-$ &  2497 &    50.6 $\pm$     7.2 &    33.5 $\pm$     4.6 & 47.8 $\pm$ 0.4\\
 $B^0$& $\mu^+\mu^-$ &   577 &     2.0 $\pm$     1.5 &     2.4 $\pm$     2.1 & 35.6 $\pm$ 0.4 \\
\hline
\end{tabular}
\label{tab:results}
\end{table}

Table~\protect\ref{tab:results} lists the efficiencies,
background composition, and number of events in the signal region based on the one-dimensional fit with a 3($\sigma$) \DeltaE requirement.
Based on Eq.(~\ref{BR_calc}) we determine:
\begin{eqnarray}
R^{+/0}(\epem)      &=& 1.019\pm 0.054 (stat)\pm  0.031(sys) \nonumber \\
R^{+/0}(\mu^+\mu^-) &=& 0.994\pm 0.049 (stat)\pm  0.033(sys) \nonumber \\
R^{+/0}(avg)        &=& 1.006\pm 0.036 (stat)\pm  0.031(sys) \nonumber
\end{eqnarray}
when assuming isospin conservation in $\B\to\jpsi K$ decays. The data sample has a mean energy of 10.580\gev\ and does not have sufficient spread to test the
hypothesis
of an energy dependent production ratio. 

We have confirmed that the result for the individual efficiency-corrected signal yields for the \jpsi\to\epem and $\jpsi\to\mu^+\mu^-$ channels 
is consistent among seven equal subsets of the full sample, as is the ratio of $\epem/\mu^+\mu^-$. 

To check our fitting technique we have performed a two-dimensional non-parametric fit to the data.  This is done by fitting the data to a sum of contributions
from
five different sources (\epem\to $\q\bar{q}$, \epem\to\ccbar, generic $\FourS\to\Bz\Bzb$, generic $\FourS\to\B^+\B^-$, and signal) whose densities~\cite{Silver}
in 
\DeltaE and \mes are determined from a non-parametric fit to candidates from Monte Carlo samples.  The two-dimensional fit is
done in the region $5.200 < \mes < 5.270$\gevcc\ and $0.030 < |\DeltaE| < 0.120$\gev.
This technique has the advantage that we are not restricted to a small range in $|\DeltaE|$.  It also employs the MC predicted background distributions, rather
than the
empirical shape imposed by the ARGUS function. The non-parametric fit method finds results that are consistent with the 
simpler counting method, both for the full sample and for data subsets.

The observed value for $R^{+/0}$ is close to one, as has been assumed by
most branching fraction measurements obtained
on the \FourS, with a ratio as large as 1.2 disfavored at the four sigma level. 
Our measurement will aid in restricting models of \FourS\ decays. 
It also allows a quantitative determination of the contribution from $R^{+/0}$ to all branching fractions that are determined at the \B factories operating on
the \FourS\ resonance.
We are grateful for the excellent luminosity and machine conditions
provided by our \pep2\ colleagues,
and for the substantial dedicated effort from
the computing organizations that support \babar.
The collaborating institutions wish to thank
SLAC for its support and kind hospitality.
This work is supported by
DOE
and NSF (USA),
NSERC (Canada),
IHEP (China),
CEA and
CNRS-IN2P3
(France),
BMBF and DFG
(Germany),
INFN (Italy),
FOM (The Netherlands),
NFR (Norway),
MIST (Russia), and
PPARC (United Kingdom).
Individuals have received support from the
A.~P.~Sloan Foundation,
Research Corporation,
and Alexander von Humboldt Foundation.


\begin{thebibliography}{99}
\bibitem{atwood} D.Atwood and W.J.Marciano, \jprd{41}, 1736 (1990).
\bibitem{lepage} G.P.Lepage, \jprd{42}, 3251 (1990).
\bibitem{manohar} R.Kaiser, A.V.Manohar, and T.Mehen, \jprl{90}, 142001 (2003).
\bibitem{voloshin} M.B.Voloshin, \mpl{A18}, 1783 (2003).
\bibitem{phi} SND Collaboration, M.N. Achasov {\em et al.}, \jprd{63}, 072002 (2001).
\bibitem{CLEO1} CLEO Collaboration, J.P.Alexander {\em et al.}, \jprl{86}, 2737 (2001).
\bibitem{CLEO2} CLEO Collaboration, S.B.Athar {\em et al.}, \jprd{66}, 052003 (2002). 
\bibitem{BABAR1} \babar\ Collaboration, B.Aubert {\em et al.}, \jprd{65}, 032001 (2002).
\bibitem{BELLE} Belle Collaboration, N.C.Hastings {\em et al.}, \jprd{67}, 052004 (2003).
\bibitem{cc} Charge conjugate decays are implied throughout this paper. Results are averages
over both charge conjugate states.
\bibitem{fleischer} R. Fleischer, and T. Mannel, \plb{506}, 311-322 (2001).
\bibitem{beamspread} \babar\ Collaboration, B.Aubert {\em et al.}, hep-ex/0308020, August 2003.
\bibitem{babar} \babar\ Collaboration, B.Aubert {\em et al.}, \nima{479}, 1 (2002).
\bibitem{FOX} G.C. Fox and S. Wolfram, \npb{149}, 413 (1979).
\bibitem{ARGUS} ARGUS Collaboration, H. Albrecht {\em et al.}, \zpc{48}, 543 (1990).
\bibitem{generic} Generic \B events include known \B decays with measured branching fractions and hadronized quark model decays for unknown branching fractions.
\bibitem{PDG2002} Particle Data Group, K. Hagiwara {\em et al.}, \jprd{66}, 010001 (2002).
\bibitem{phasesp} The phase space difference between these two decays is negligible.  
\bibitem{Silver} ``Density Estimation'', B. W. Silverman, CRC Press, ISBN: 0412246201 (1986).
\end{thebibliography}
\end{document}